# Flexible plasma multi-jet source operated in radial discharge configuration

Carles Corbella[1,a)] and Sabine Portal[1]

[1] Department of Mechanical & Aerospace Engineering, George Washington University, 800 22nd Street, Northwest, Washington, DC 20052, United States of America

Following up a recent study describing a flexible plasma source operated in planar geometry, the performance of a cold atmospheric plasma jet (CAPJ) matrix emanating radially from a soft cylindrical surface in the open air is presented. The plasma device, which has a set of small outlets produced in its side surface, has a length of 12 cm and an outer diameter of 5.4 cm. The dielectric barrier discharge (DBD) sustaining the plasma jets is generated in helium flowing between two coaxial electrodes, which are separated by both an insulating tape and a cylindrical wall made of polymer foam. Two operation modes are considered: four equidistant CAPJs at the same axial position (brush mode) and three aligned CAPJs at constant azimuthal angle (comb mode). All discharges, excited by 15 kHz-AC voltages at 3.8 kV in amplitude, have resulted in uniform lengths and intensities of the jets. Consumed discharge powers of between 0.1 and 1 W have been estimated from current-voltage measurements. Optical emission spectroscopy (OES) has shown the preponderance of hydroxyl groups, nitrogen molecules and helium atoms in the active DBD region and the jet afterglow zone. This new design of CAPJ anticipates promising applications for treating the inner surface of hollow and delicate components for activation or healing purposes.

The need in plasma medicine for extended and soft plasma sources, basically in the form of dielectric barrier discharges (DBD), able to access or treat complex topologies has been an active focus of research for different applications.[1-7] Followed by this motivation, Corbella et al. have recently characterized a flexible platform of cold atmospheric plasma jet (CAPJ) arrays consisting of a flat and bendable aerogel matrix.[8] Such a plasma source can be extremely helpful in healing surgical margins in cancer treatment. This note extends the previous study after modifying the discharge geometry of the original device. In particular, the main characteristics of a radial matrix of CAPJs generated in a soft cylindrical polymer are shown. The basic structure of the plasma source prototype, as well as the electrical and optical properties of the DBD, are discussed.

The new plasma source consists of a DBD active zone and an afterglow zone (Fig. 1). The latter constitutes the propagation region of CAPJs. The active region occupies the space between two coaxial electrodes separated by a hollow cylinder, of 12 cm length, made of polyethylene (PE) foam. This material was selected due to its flexibility and optimal gas barrier properties. The inner and outer diameters are 2.5 cm and 5.4 cm, respectively. The conducting core (biased electrode) is an aluminum rod of 3 mm in diameter. It was covered with Kapton tape, which acts as a dielectric barrier. The external electrodes (grounded electrodes) consist of copper tape adhered onto the external PE housing surface around its external circumference. All outer electrodes were separated 15 mm each other. The CAPJs were flowing through the PE spacer by means of channels of around 1 mm in diameter. The discharges were ignited with He (99.995%), which was introduced into the PE cylinder at one end.

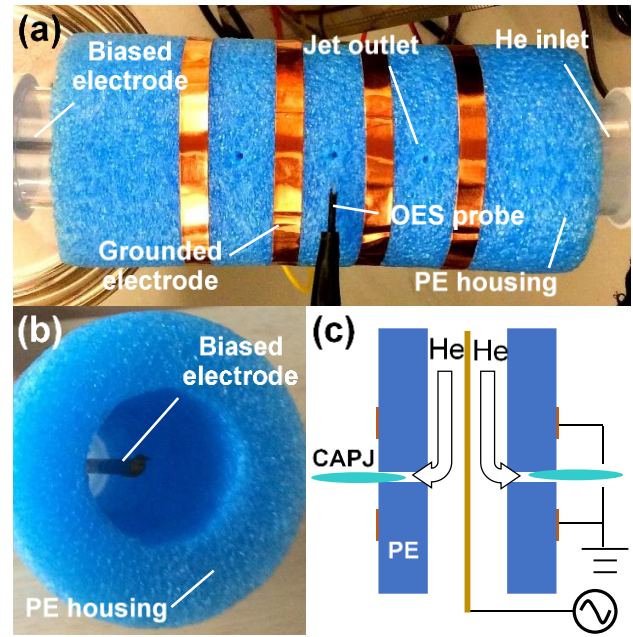

FIG. 1. (a) Image of the cylindrical plasma source showing the different parts. (b) Detail of the DBD active zone, comprised between the biased (core) electrode and the flexible PE spacer. (c) Schematic cross section of the plasma source. The working principle of the radial CAPJ generation in the afterglow zone is indicated.

The electrical powering circuit and diagnostics are described elsewhere.[8] In brief, the supplied DC power is modulated with the sinusoidal voltage from a function

a) Author to whom correspondence should be addressed. Electronic mail: ccorberoc@gwu.edu.

generator. The resulting waveform is transformed into a high voltage AC signal, which is applied to the core electrode. The discharge parameters were similar to those used in our previous study.[8] In the present cylindrical geometry, the DBD region was fed with an AC voltage of 7.5 kV peak to peak at a frequency of 15 kHz. The inlet He flow rate was adjusted to a He rate per outlet of 2 slm, so that the CAPJs could flow in or near laminar regime.[9] Note that the flow rate per CAPJ is estimated as the total He rate divided by the number of outlets.

As shown in Fig. 2, two exclusive operation modes providing CAPJs with a laminar pattern were considered: (1) *brush mode*, in which four equidistant CAPJs are produced around the PE surface at the middle axial position (azimuth angles: 0º, 90º, 180º and 270º); and (2) *comb mode*, where CAPJs emanated from three outlets, 20 mm apart each other, aligned on the axial direction. The jets produced in both modes exhibited stationary and uniform plasma plumes with a length of the order of 1 cm. Such a homogeneity in CAPJ appearance can be explained by the axisymmetric geometry of the source (brush mode) and the periodicity of electric field strength at the outlet positions along the axial direction (comb mode). The electric field distributions were computed from simulations of the electrostatic potential performed by finite element method (FEM) approach (not shown here).

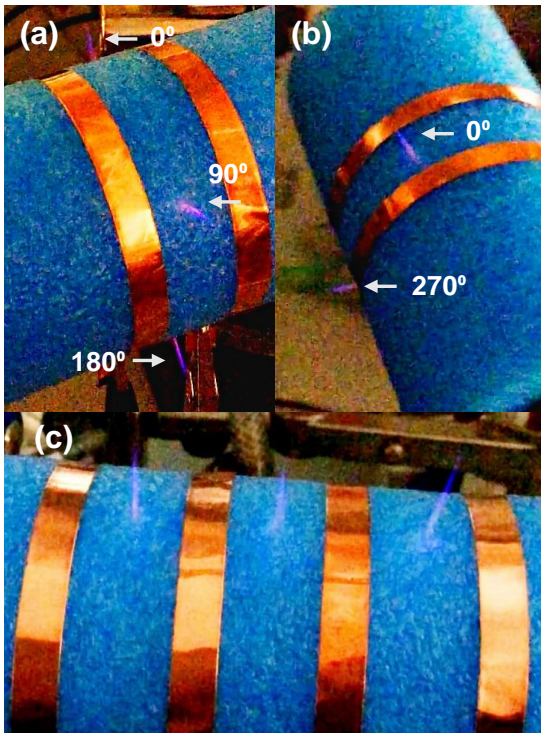

FIG. 2. Images of the He CAPJs striking in the (a,b) brush mode (8 slm He) and (c) the comb mode (6 slm He). The jet positions at 0º, 90º, 180º and 270º in (a) and (b) are indicated with arrows. Two grounded electrodes were used in the brush mode, whereas four grounded electrodes were necessary to obtain CAPJs in the comb mode.

Fig. 3 shows waveforms of the applied AC voltage and current for the brush and comb mode operations. The discharge current was estimated by subtracting the displacement current (no He flow) from the total current measured when the CAPJs were striking (with He flow) at a constant voltage amplitude of 3.8 kV. The discharge currents exhibited asymmetric, periodical profiles composed of current peaks and a sinusoidal part. The current values were comprised between -4 mA and 3 mA in the brush configuration, and -5 mA and 4 mA in the comb mode. The consumed power in each case was calculated by averaging the product of current and voltage within one AC period. From the obtained curves, approximated electrical powers of 0.1 W for the comb mode and 1 W for the brush mode were estimated.

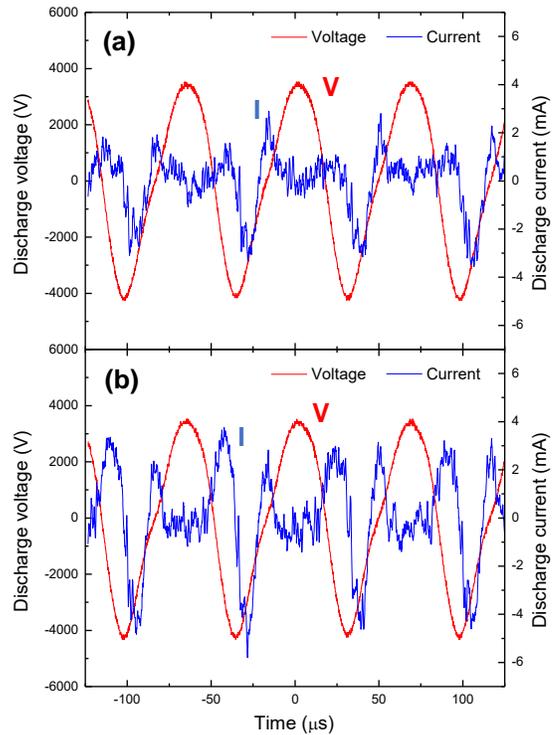

FIG. 3. Current (blue) and voltage (red) waveforms corresponding to He CAPJs ignited in (a) brush mode at 8 slm and (b) comb mode at 6 slm.

The current profiles reflected a characteristic peak occurring during the initial streamer propagation in the positive phase, which is typical in DBD generation.[10] On the other hand, the sinusoidal component might be originated by dissipation effects in the discharge region. The time shift towards the right with respect to voltage observed in the current curves is associated to a dominant inductive component in the equivalent circuit of the discharge. Further discussion on the physical basis of the waveforms can be found in our previous study.[8] Finally, the higher values in power measured when operating in brush mode (≈1 W) vs. comb mode (≈0.1 W) are probably connected to an increased resistivity at the basis of the four CAPJs, which share the same axial position. This hypothesis can be investigated by means of fluid simulations considering the balance of mass and energy fluxes incident to the plasma volume.

OES analysis of the plasma jets was carried out using a StellarNet spectrometer operating at a spectral range 191.0-851.5 nm with a spectral resolution of 0.5 nm. The UV-visible spectra were recorded with an integration time of 2 s. Only the emitted spectra corresponding to the plasma source in comb mode were measured. In this configuration, the OES probe was either inserted into the central outlet to explore the active region (Fig. 4a), or it was placed outside the source to study the optical emission from the afterglow region (CAPJ) expanding into the atmosphere (Fig. 4b). In the afterglow analysis, the axial and radial positions of the OES probe were both 5 mm.

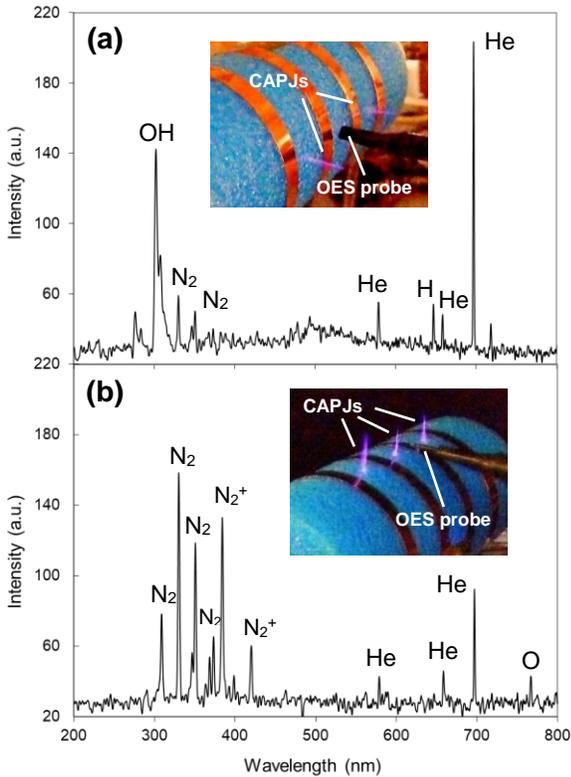

FIG. 4. Optical emission spectra from the He CAPJs measured (a) in the active DBD zone and (b) in the afterglow zone. The optically active plasma species are identified. Total He rate was 4 slm. Insets: images of the OES probe position with respect to the CAPJs.

The emission lines have been assigned to the presence of $N_2$, $N_2^+$, He, O, H and OH species, whose transitions have been reported elsewhere.[8,11,12] The spectrum emitted by the CAPJ (afterglow) shows a profile similar to that observed in the flat source.[8] In fact, the strong nitrogen emission is consistent with a free He plasma mixed with air from the atmosphere. Interestingly, the spectrum measured in the active zone is substantially different: besides He, significant emission from OH groups and small contributions by $N_2$ and H show up and are attributed to gas impurities (air and water).[11]

Finally, it is noteworthy to mention that the new plasma source presented here stays at room temperature after several minutes of operation. In addition, the source is stable under moderate deformation (lateral compression) of the PE housing. In conclusion, the performance of a new plasma source based on CAPJs emanating radially from a soft polymeric cylinder has been briefly characterized. A more detailed report, including source bending capabilities and control of each jet separately, is planned for a future paper. The applications envisaged for this source include plasma cleaning and healing of cavities in delicate samples, like organic tissues and temperature-sensitive materials. This study has proven that flexible multi-jet prototypes provide a robust platform for fabricating plasma devices with the jets oriented in both axial and radial directions to uniformly treat samples with virtually any topography.

This work was supported by National Science Foundation through the award nr. 1919019. The authors are grateful to Prof. M. Keidar and Dr. L. Lin (George Washington University) for scientific advising.